\documentclass[conference]{IEEEtran}

\usepackage{xcolor}
\usepackage{booktabs}
\usepackage{multirow}
\usepackage[utf8]{inputenc}
\usepackage{amsmath, amssymb}
\usepackage{graphicx}
\usepackage{cite}
\usepackage{geometry}
\usepackage[mathscr]{euscript}
\usepackage{multirow}
\begin{document}

\title{Data driven approach for Outdoor Channel Prediction in 5G and Beyond}
\author{\IEEEauthorblockN{A. Sathi Babu}
\IEEEauthorblockA{\textit{Department of ECE } \\
\textit{SRM University AP}\\
Guntur, India \\
sathibabu\_arepalli@srmap.edu.in }\\
\and
\IEEEauthorblockN{V. Udaya Sankar }
\IEEEauthorblockA{\textit{Department of ECE } \\
\textit{SRM University AP}\\
Guntur, India \\
udayasankar.v@srmap.edu.in }\\
\and
\IEEEauthorblockN{Vishnu Ram OV}
\IEEEauthorblockA{\textit{Independent Research Consultant} \\
Bengaluru, India \\
vishnu.n@ieee.org}
}
\date{}
\maketitle

\begin{abstract}

An evolution of Wireless Communications towards 5G and beyond provides improved user experience in terms of quality of services. Understanding and estimating Channel information plays crucial role in providing better user experience. Traditional methods of channel estimation involves periodically sending pilots (known signals), estimating channel and send back estimated channel information to the BS  which increases computational complexity and communication complexity. Hence, we focus on data driven approach for channel estimation. In this work, we explore a channel estimation mechanism at 7GHz frequency band for a given user location. This work involves data generation using Ray tracing mechanism and Machine learning model training that contains feature variables such as transmitter location, user location and target variable as channel coefficient . We explored Support Vector Regression, K-nearest neighbor (KNN), Random Forest, XGBoost and MLP. We found via simulations that XG Boost and proposed MLP performs better than Support Vector Regression, KNN and Random forest regression. 
\end{abstract}

\section{Introduction}

Evaluation of Wireless Communications (WC) from 1G to 5G and beyond satisfies Quality of service (QoS) of users such as high data rate, low latency and reliable communication. Accurate channel information at Transmitter (Tx) or Receiver (Rx) is crucial for satisfying QoS of all users. In traditional wireless communication systems (WCS), Rx estimates channel using pilot symbols from Tx in a specific periods of time (known as periodic channel estimation). As number of subchannel/subcarrier increases, computation complexity and communication complexity also increases. This motivates towards data driven approaches for channel estimation. An overview of channel estimation methods both traditional and dat driven approaches are studied in \cite{li2025survey}.

An overview of challenges and opportunities in  the 7-24 GHz band is studied in \cite{cui20256g}. Traditional stochastic channel models provide statistical representations of wireless propagation but often fail to capture site-specific propagation characteristics, particularly in complex indoor and outdoor environments. Deterministic ray-tracing models address this limitation by modelling the physical propagation trajectories based on the geometrical and material properties of the surrounding environment\cite{mbugua2020review}. 

A practical ray tracing (RT) models to characterize the propagation channels are studied in \cite{7152831}. The RT methods are computationally expensive for large-scale simulations and require precise environmental information. This motivates towards exploring methods such as combine ray-tracing simulations with machine learning (ML) techniques. In this framework,RT is used to generate physically consistent datasets that capture realistic propagation behaviour and Machine learning (ML) models are used to learn spatial-channel relationships from these datasets \cite{li2025deeprt}. An efficient prediction of channel state information (CSI) using ML is proposed in \cite{lee2023lightweight}. In \cite{alkhateeb2019deepmimo}, authors generated DeepMIMO dataset that can be used for ML approach towards predeiction of channel characteristics. An ML based approch towards estimation of real time channel statistics and propogation characteristics of physical environments is studied in \cite{saeizadeh2024digitaltwin}. Also, in \cite{hoydis2018deep}, authors studied channel estimation using Deep Learning (DL) approach. 

An RT generated synthetic datasets are used to train ML models like Random Forest (RF) and K-nearest neighbour (KNN) to predict path loss (PL) in suburban environments \cite{tarhouni2025machine}. Where as in  \cite{inproceedings}, an RT based approach that utilizes full environment geometry, material properties is proposed with Tx-Rx positions as input features to ML models to predict the path loss. An RT generated datasets are used to predict channel parameters like Angle of Arrival (AoA), Angle of Departure (AoD) and Receiver Signal Strength (RSS) using neural network approaches in studied in \cite{8422221}. A Deep learning ResNet-based path loss modeling approach that uses environment-aware features including relative height maps and distance maps (derived from TX-Rx geometry) is studied in \cite{Nagao2022}. Motivated by these works, in this paper, we focus on prediction of baseband complex channel coefficient at Base Station (BS) given location of User Equipment (UE). We used RT model to generate data set and ML approach to predict channel coefficient.  

%  A Geometric based channel prediction methods that utilizes environment maps and user positions to predict the CSI  is provided in \cite{deutschmann2025geometry}. 

The remainder of this paper is organized as follows. Section II presents System model, Section III describes Solution Approach, Section IV Provides Simulation results and analysis and finally we concluded in the Section V. 

\section{System model}

\begin{figure}[htbp]
    \centering
    \includegraphics[width=0.9\linewidth]{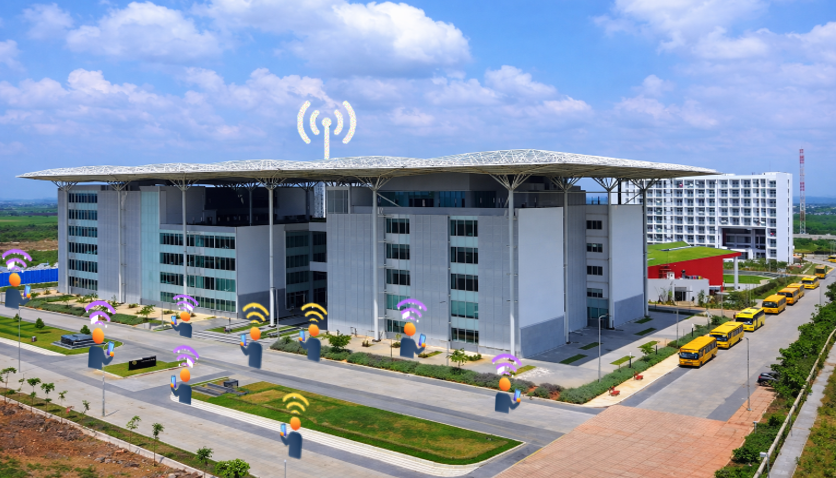}
    \caption{SRM University AP, India, an outdoor campus scenario,with rooftop base station and multiple UEs}
    \label{fig:srm_outdoor_scenario}
\end{figure}

We consider a small base station (SBS) with single antenna mounted on a top of building and it serves a set of users (UEs) with in a small specific region (Fig. \ref{fig:srm_outdoor_scenario}). We assume Line of sight (LOS) points (UEs) from SBS, but there exists both LOS and Non LOS paths between SBS and UEs. We consider downlink (DL) single carrier transmissions in the narrow band channel.

Let $x$ be the transmitted symbol, $h$ be the baseband complex wireless channel coefficient ($h=h_{re}+jh_{im}$) and $y$ be received symbol, then
\begin{equation*}
     y = hx+n
\end{equation*}
where, $n$ is complex additive white gaussian noise (AWGN). For reliable communication, we need best channel estimation $\hat{h}$ such that $\hat{h} \approx h$. Traditionally, channel is estimated by UE using pilots or known symbols and send back this channel information to BS. In general this process continues periodically and it leads to increased computational complexity at UE and increased overall communication complexity between BS and UE. 

In \cite{hoydis2021toward}, the authors explained the benefits of replacing traditional blocks in the wireless communication chain using AI enabled blocks. One of the important block is joint channel estimation and equalization. In this work we focus on data driven approach for channel prediction.

\section{Solution Approach}

Our goal is to predict the channel $h$ for the UE when BS knows the location of UE. In this context, we assume that UE location is available at the BS. We use the following steps. 
\begin{enumerate}
     \item Generate the channel coefficient $h$ for a given set of Tx and Rx location pair for the fixed frequency of transmission
     \item Channel prediction using Machine Learning (ML) model from generated data 
\end{enumerate}
    
\subsection{Data (Channel coefficient) Generation}

\begin{figure}[htbp]
    \centering
    \includegraphics[width=0.9\linewidth]{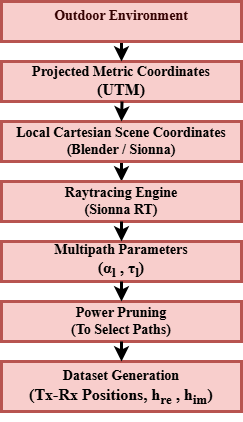}
    \caption{Overview of Dataset Generation}
    \label{fig:blockdiagram}
\end{figure}

An overview of data generation flow is given in Fig. \ref{fig:blockdiagram}, receiver coordinate points from outdoor environment is first converted into UTM coordinates. This UTM coordnates is converted into local coordinates using Blender/Sionna as Sionna Ray Tracing (RT) model works only with local coordinates. 

The multipath parameters $(\alpha_i,\tau_i)$ are generated using Sianna RT, where $\alpha_i$ is path gain and $\tau_i$ is the path delay of path $i$. Let $h(\tau)$ be multipath channel impulse response then, 

\begin{equation}
h(\tau)=\sum_{i \in \mathscr{N}}\alpha_i\,\delta(\tau-\tau_i)
\end{equation}

Where, $\mathscr{N}$ is a set of all multipath (LOS \& NLOS) components. At higher frequencies, the wireless channel exhibits sparsity where only a few dominant paths carry most of the signal energy \cite{misra2015millimeter}. There are some paths that have insignificant power which can be ignored. We use a power-based pruning strategy \cite{9501275} as described. Let $P_i$ be the power of $i^{th}$ multipath component, then

\begin{equation}
P_i = |\alpha_i|^2
\end{equation}
and the maximum receiving power among all multipath components is given by

\begin{equation}
P_{\max} = \max_i |\alpha_i|^2
\end{equation}
Let $\Delta_i$ (in dB) be the power ratio between strongest path (that carriers maximum power) and $i^{th}$ multipath component is given by 
\begin{equation}
\Delta_{\mathrm{i}} = 10 \log_{10} \left( \frac{P_{\max}}{P_i} \right)
\end{equation}

We consider a predefined relative power threshold $\Delta_{th}$ (in dB) to  eliminate insignificant paths and a multipath component is retained if
\begin{equation*}
\Delta_i \leq \Delta_{\mathrm{th}}    
\end{equation*}

which, implies 

\begin{equation}
P_i \geq P_{\max} \cdot 10^{-\Delta_{\mathrm{th}}/10}.
\end{equation}

We use normalized transmit power $P_{TH}$ and for a particular carrier frequency $f_c$, the narrow band complex channel coefficient corresponding to the channel impulse response (CIR) is 

\begin{equation}
h = \sum_{i \in \mathscr{N}_{th}} \alpha_i e^{-j2\pi f_c \tau_i}
\end{equation}

Where, $\mathscr{N}_{th}$ is a set of multipath components such that $\mathscr{N}_{th} = \{i: \left| \tau_i - \frac{d_i}{c} \right| \leq \epsilon_{\tau} \} $ with $\epsilon_{\tau}$ is a small value depending on the environment/scene. The resulting data has the following features $(P^{Tx},P^{Rx},h_{re},h_{im})$ where $P^{Tx} = (P^{Tx}_x,P^{Tx}_y,P^{Tx}_z)$ local transmitter coordinates, $P^{Rx}=(P^{Rx}_x,P^{Rx}_y,P^{Rx}_z)$ local receiver coordinates, $\bar{h} = (h_{re},h_{im})$ are real and imaginary part of baseband channel coefficient. A structured dataset generated from the Sionna ray tracing engine is given as
\begin{equation*}
    \mathscr{D} = \{(P^{Tx,l},P^{Rx,l},h^l_{re},h^l_{im})\}_{l=1}^L \sim iid \quad \mathscr{P}_{X,\bar{h}}
\end{equation*}
where, $X^l = (P^{Tx,l},P^{Rx,l})$ is $l^{th}$ feature and $\bar{h}^{l} = (h^l_{re},h^l_{im})$ is $l^{th}$ target variable. 

\subsection{Channel prediction using ML model}

An elaborated discussion on development of machine learning algorithms using probabilistic approach is studied in \cite{murphy2022probabilistic} and also Regression models are widely used for channel modeling and channel prediction strategies\cite{electronics12234740}. We consider $\mathscr{D}_{re} = \{(P^{Tx,l},P^{Rx,l},h^l_{re})\}_{l=1}^L \sim iid \quad \mathscr{P}_{X,\bar{h}_{re}}$  and $\mathscr{D}_{im} = \{(P^{Tx,l},P^{Rx,l},h^l_{im})\}_{l=1}^L \sim iid \quad \mathscr{P}_{X,\bar{h}_{im}}$ for ML model training. We use two seperate regression models since our goal is to estimate $P_{\frac{\bar{h}_{re}}{X}}$ and $P_{\frac{\bar{h}_{im}}{X}}$ which are independent but are not identical. An ML based workflow for channel prediction is given in Fig. \ref{fig:ml_workflow}. 
\begin{figure}[htbp]
\centering
\includegraphics[width=\columnwidth]{ml_model_final.pdf}
\caption{Model workflow for channel prediction}
\label{fig:ml_workflow}
\end{figure}

We consider data set $\mathscr{D}_{re} = \{(P^{Tx,l},P^{Rx,l},h^l_{re})\}_{l=1}^L$ and we devide dataset into train set and test set using two approachs as given below.
\begin{enumerate}
    \item \emph{Random Sampling}: The complete dataset is randomly divided into 80\% training and 20\% testing samples using fixed random seed. 
    
    \item \emph{Spatial Sampling}: We group receiver positoins into varous spatial blocks for testing and remaining spatial blocks are used for train the model. We use this process to avoid spatial leakage and provide a strict assessment of model generalization to unseen spatial regions\cite{bolatbek2026physically}.     
\end{enumerate}

We consider Support Vector Regression (SVR) with RBF kernal and K-Nearest Neighbour (KNN) regression, Random Forest, XG-Boost regression as data might be non-linearly separable. We normalize the train set using standardization before train the model. This ensures improvement in the performance of the model since $P_{\frac{\bar{h}_{re}}{X}}$ has normal distribution. Accordingly we normalize the remaining sets before testing performance of the model. Also, we considred the Multilayer Perceptron (MLP) model for training and testing with Min-Max scaler for dataset. As illustrated in Fig.~\ref{fig:MLP_FINAL}, MLP consists of fully connected hidden layers with the structure containing 128, 64, and 32 neurons. A ReLU activation function is applied for hidden layers and the output layer is linear. 

We consider Root mean square error (RMSE), Mean absolute error (MAE), $R^2$ and Adjusted $R^2$ metrics for evaluating the model performance as described below. Let $n$ be the number of test samples, $y_i$ be the target value and $\hat{y}_i$ be the predicted target value, then \\
\textbf{Root Mean Square Error (RMSE)}  is given as 
\begin{equation}
RMSE = \sqrt{\frac{1}{n}RSS}
\end{equation}
Where, residual sum of squares $RSS = \sum_{i=1}^n (y_i - \hat{y}_i)^2$. Lower the $RSS$ better the model fit. \\
\textbf{Mean Absolute Error (MAE)} quantifies the average magnitude of the absolute difference between predicted and actual channel coefficients and is given by
\begin{equation}
MAE = \frac{1}{n}\sum_{i=1}^{n} |y_i - \hat{y}_i|
\end{equation}

\textbf{Coefficient of Determination ($R^2$)} measures the proportion of variance in the observed data:
\begin{equation}
R^2 = 1 - \frac{\sum_{i=1}^n (y_i - \hat{y}_i)^2}{\sum_{i=1}^n (y_i - \bar{y})^2} = 1 - \frac{RSS}{TSS}
\end{equation}
where, $\bar{y} = \frac{1}{n}\sum_{i=1}^n y_i$, total sum of squares $TSS = {\sum_{i=1}^n (y_i - \bar{y})^2}$. Larger the value of $R^2$ implies better the model fit to the data. \\
\textbf{Adjusted coefficient of determination (Adjusted $R^2$)} accounts for number of input input features ($k$) and the number of observations/test samples ($n$) and given as 
\begin{equation}
R_{\mathrm{Adj}}^{2}
=
1-
\frac{(1-R^2)(n-1)}
{n-k-1}
\end{equation}

The data $\mathscr{D} = \{(P^{Tx,l},P^{Rx,l},h^l_{re},h^l_{im})\}_{l=1}^L$ is generated at a large frequency $fc$ ($\approx 7GHz$) and a change in distance ($\frac{\lambda}{2} \approx 2.14cm$) causes $\pi$-radian phase shift in corresponding complex channel coefficient $h$. Hence, a small change in receiver location for a fixed transmitter location causes large variations in values of $h^l_{re},h^l_{im}$ that leads poor model fitting (which we see in results). We use phase compensation before model training and testing as described below. 

Let $h^l = h^l_{re}+jh^l_{im}$, then $g^l = h^le^{j2\pi f_c d_l/c} = g^l_{re}+jg^l_{im}$ where $d_l$ is euclidean distance between transmitter and $l^{th}$ receiver location given as $d_l = ||P^{Tx,l}-P^{Rx,l}||_2$. We use $\mathscr{D} = \left\{\left(P^{Tx,l}, P^{Rx,l}, g^l_{re}\right)\right\}_{l=1}^{L}$ for model training and testing. It is easy to reconstruct $\hat{h}^l$ from predeicted $\hat{g}^l$ using $\hat{h}^l = \hat{g}^le^{-j2\pi f_c d_l/c}$.

We use a similar approach for another data set $\mathscr{D}_{im} = \{(P^{Tx,l},P^{Rx,l},h^l_{im})\}_{l=1}^L$ for model training and model recommendation.   

\begin{figure}[htbp]
\centering
\includegraphics[width=\columnwidth=0.8]{MLP_FINAL.pdf}
\caption{An MLP model}
\label{fig:MLP_FINAL}
\end{figure}

\section{Simulation Results and Analysis}

We use a framework that integrates (i) physically consistent 3D environment modeling in Blender with detailed geometry and material assignments and (ii) a deterministic electromagnetic ray tracer implemented using NVIDIA Sionna RT for data generation. Each data point from the dataset corresponds to a fixed transmitter (Tx) and receiver (Rx) location configuration, enables a spatially indexed channel representation that is suitable for training ML model for channel prediction.

We consider a $0.3 \times 0.3$ km radius outdoor campus environment (Fig. \ref{fig:srm_outdoor_scenario}) located beside the Administrative Block of SRM University within latitude and longitude ranges of 16.46269–-16.46564, 80.50635-–80.50887. The building footprints were imported from OpenStreetMap data and a realistic building heights consistent with urban morphology were maintained. Buildings, roads, vegetation, and terrain are included to maintain spatial realism. The material properties are choosen to approximate electromagnetic interactions at 7\,GHz since, material configuration enables ray tracing simulation to capture geometric-based multi-path propagation effects. The balance between specular and diffuse reflections are controlled by surface roughness, where as the reflection and transmission characteristics are governed by the Fresnel equations. 

We consider the following parameter for our simulation experiments as given in Table \ref{tab:simulation_parameters_outdoor}. 

\begin{table}[h]
\caption{Simulation Configuration for Outdoor Scenario}
\label{tab:simulation_parameters_outdoor}
\centering
\begin{tabular}{lc}
\hline
\textbf{Parameter} & \textbf{Value} \\
\hline
Carrier Frequency & 7\,GHz \\

Environment Size & $0.3 \times 0.3$ km \\

Transmitter Height & 21.44 m \\

Receiver Height & 1.5 m \\

Antenna Type & Isotropic \\

Transmitter Antenna Array & Planar Array $(1 \times 1)$ \\

Receiver Antenna Array & Planar Array $(1 \times 1)$ \\

Antenna Polarization & Vertical \\
 
$\Delta_{\mathrm{th}}$ & $30\,\mathrm{dB}$\\
$P_{TH}$ & $1W$ \\

$\epsilon_{\tau}$ & $57.76ns$ \\

Random Seed & 42 \\
\hline
\end{tabular}
\end{table}

\begin{figure}[htbp]
    \centering
    \includegraphics[width=0.9\linewidth]{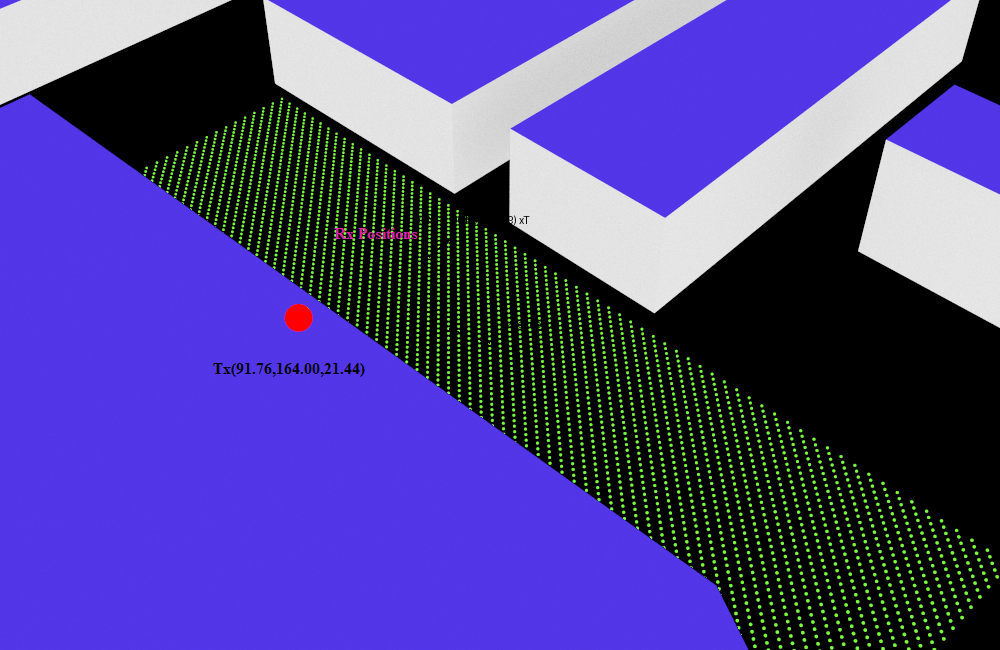}
    \caption{SRM University-AP, India, with a fixed Tx position and 10,000 receiver locations at coordinates of 16°4649N, 80°5078W.}
    \label{fig:srm_outdoor_model}
\end{figure}

We consider the Sionna Ray-tracing framework \cite{aoudia2025sionnart} to simulate multipath propagation between the transmitter and receiver locations. A fixed transmitter is considered while the receiver positions are  uniformly distributed across the environment (Fig. \ref{fig:srm_outdoor_model}) to capture the spatial channel variations. We approximately considered 10000 receiver positions across the given coordinates. We considered receiver height as 1.5\.m according to the 3GPP channel modeling recommendations \cite{3gpp38901R17}.

A snapshot of the structured data set $\mathscr{D} = \{(P^{Tx,l},P^{Rx,l},h^l_{re},h^l_{im})\}_{l=1}^L$ for a fixed Tx position (m): (91.76, 164.00, 21.44) is shown in Table \ref{tab:unscaled_h_components}. 

\begin{table}[htbp]
\centering
\caption{Representative samples of channel components $h_{\mathrm{re}}$ and $h_{\mathrm{im}}$}
\label{tab:unscaled_h_components}
\resizebox{\columnwidth}{!}{%
\begin{tabular}{c c c c}
\hline
\textbf{Sample} &
\textbf{Rx Position (m)} &
$\mathbf{h_{\mathrm{re}}}$ &
$\mathbf{h_{\mathrm{im}}}$ \\
\hline

1 &
$(29.896606,\ 120.701283,\ 1.50)$ &
$-2.21568019 \times 10^{-7}$ &
$6.42114636 \times 10^{-5}$ \\

2 &
$(31.252259,\ 120.697978,\ 1.50)$ &
$1.28577282 \times 10^{-5}$ &
$8.41804226 \times 10^{-6}$ \\

3 &
$(32.607911,\ 120.694674,\ 1.50)$ &
$3.96517681 \times 10^{-6}$ &
$-5.55413541 \times 10^{-5}$ \\

\hline
\end{tabular}%
}
\end{table}

\begin{figure}[h]
\centering
\includegraphics[width=\linewidth]{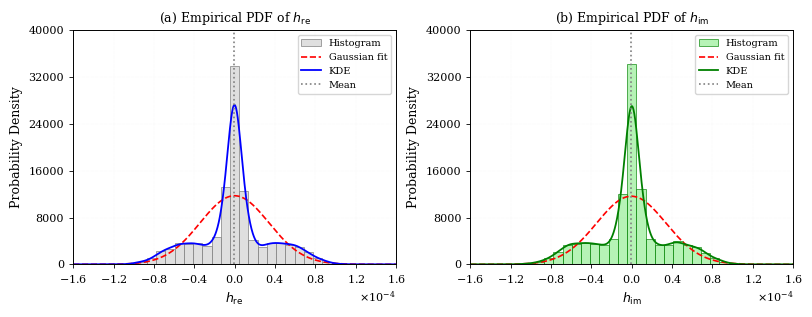}
\caption{PDF Distributions of channel components $h_{\mathrm{re}}$ and $h_{\mathrm{im}}$}
\label{fig:mean_plots_without_weight}
\end{figure}

The probability density functions (PDFs) of real and imaginary components ($h_{re},h_{im}$) of the channel coefficients are given in Fig. \ref{fig:mean_plots_without_weight} and its statistical properties (mean, variance) are given in Table \ref{tab:unscaled_h_statistics}.

\begin{table}[htbp]
\centering
\caption{Statistical Properties of channel components $h_{\mathrm{re}}$ and $h_{\mathrm{im}}$}
\label{tab:unscaled_h_statistics}

\resizebox{\columnwidth}{!}{%
\begin{tabular}{c c c c}
\hline
\textbf{Component} &
\textbf{Mean} &
\textbf{Variance} &
\textbf{Std. Dev.} \\
\hline

$\mathbf{h_{\mathrm{re}}}$ &
$-1.489595 \times 10^{-7}$ &
$1.157963 \times 10^{-9}$ &
$3.402886 \times 10^{-5}$ \\

$\mathbf{h_{\mathrm{im}}}$ &
$-4.831864 \times 10^{-7}$ &
$1.174364 \times 10^{-9}$ &
$3.426900 \times 10^{-5}$ \\

\hline
\end{tabular}%
}
\end{table}

It is observed that PDFs of $h_{re}, h_{im}$ are normal with almost same variance but has different mean values. Hence, we go for different regression models for prediction of channel coeffcients. The spread of distribution indicates that has reduced multipath richness which is consistent with dominated line-of-sight propagation and a limited number of scattering paths.

A subset of randomly selected 8000 samples from 10000 for model training (80\%) and the remaining unseen samples (2000) are used as an independent test dataset to evaluate generalization capability of the trained models. We standardize the features of the generated dataset to improve performance of the ML model using $\tilde{x}_{i,k} = \frac{x_{i,k} - \mu_k}{\sigma_k}$, for $k^{th}$ feature of $i^{th}$ data and for the MLP use considered Min-max scaling. The model training parameters for MLP is shown in Table \ref{tab:model_training_mlp}.

\begin{table}[h]
\caption{Model training parameters for MLP}
\label{tab:model_training_mlp}
\centering
\begin{tabular}{lc}
\hline
\textbf{Parameter} & \textbf{Value} \\
\hline
Loss function & Mean Square Error (MSE) \\

Optimizer & Adam \\

Learning rate ($\eta$) & 0.001 \\

Batch Size & 64 \\

Epochs & 500\\

\hline
\end{tabular}
\end{table}

\begin{figure}[h]
\centering
\includegraphics[width=\columnwidth]{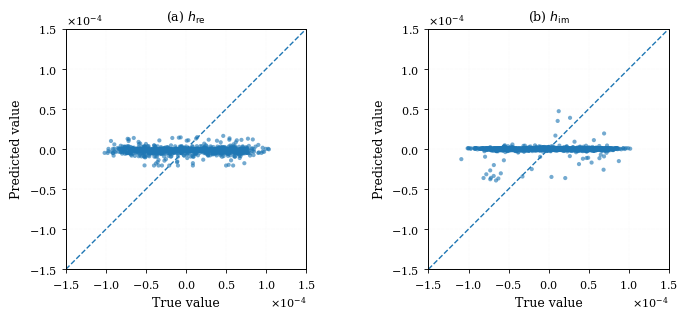}
\caption{Actual versus predicted channel coefficients under the random split using MLP
(a) $h_{\mathrm{re}}$ and (b) $h_{\mathrm{im}}$.}
\label{fig:direct_random_mlp}
\end{figure}

Where as for the spatial sampling, we generated 8108 training samples and 1892 test samples by considering 6 spatial blocks. Here also, we considered standard scaling for ML models and Min-max scaling for MLP model. The performance of the trained models on the test data is shown in Table ~\ref{tab:unscaled_direct}. It is observed from the table that all models give smaller prediction errors but these models does not fit data well due to poor or negative $R^2$ value. This is evident from figure Fig. \ref{fig:direct_random_mlp} for random sampling that describes relation between actual and predicted channel coefficients is not diagonal \cite{murphy2022probabilistic} and we observed similar kind of results for spatial sampling. We now look at the phase compensated dataset as described below. 

\begin{table*}[!t]
\centering
\caption{Model performance for $h_{\mathrm{re}}$ and $h_{\mathrm{im}}$}
\label{tab:unscaled_direct}

\footnotesize
\renewcommand{\arraystretch}{1.08}
\setlength{\tabcolsep}{4pt}

\begin{tabular*}{\textwidth}
{@{\extracolsep{\fill}}cclcccc@{}}

\toprule
\textbf{Split} &
\textbf{Target} &
\textbf{Model} &
$\mathbf{R^2}$ &
\textbf{Adjusted }$\mathbf{R^2}$ &
\textbf{MAE} &
\textbf{RMSE} \\
\midrule

\multirow{10}{*}{Random}
& \multirow{5}{*}{$h_{\mathrm{re}}$}
& SVR(RBF)
& -0.0169
& -0.0200
& 0.000023553
& 0.000034299 \\

&
& KNN
& -0.1442
& -0.1477
& 0.000024586
& 0.000036382 \\

&
& Random Forest
& -0.2048
& -0.2084
& 0.000024413
& 0.000037333 \\

&
& XGBoost
& -0.0435
& -0.0466
& 0.000023884
& 0.000034743 \\

&
& MLP
& -0.0165
& -0.0196
& 0.000023697
& 0.000034292 \\

\cmidrule(lr){2-7}

& \multirow{5}{*}{$h_{\mathrm{im}}$}
& SVR(RBF)
& -0.0203
& -0.0234
& 0.000024961
& 0.000036057 \\

&
& KNN
& -0.1115
& -0.1149
& 0.000025546
& 0.000037634 \\

&
& Random Forest
& -0.1636
& -0.1672
& 0.000025292
& 0.000038506 \\

&
& XGBoost
& -0.0322
& -0.0354
& 0.000024981
& 0.000036267 \\

&
& MLP
& 0.0026
& -0.0004
& 0.000024736
& 0.000035649 \\

\midrule

\multirow{4}{*}{Spatial}
& \multirow{2}{*}{$h_{\mathrm{re}}$}
& XGBoost
& -0.0245
& -0.0278
& 0.000016605
& 0.000028158 \\

&
& MLP
& -0.0122
& -0.0154
& 0.000016447
& 0.000027988 \\

\cmidrule(lr){2-7}

& \multirow{2}{*}{$h_{\mathrm{im}}$}
& XGBoost
& -0.0816
& -0.0850
& 0.000016860
& 0.000028957 \\

&
& MLP
& -0.0020
& -0.0051
& 0.000016247
& 0.000027871 \\

\bottomrule

\end{tabular*}

\end{table*}

%\begin{figure}[h]
%\centering
%\includegraphics[width=\columnwidth]{direct_spatial.png}
%\caption{\textcolor{red}{
%Actual versus predicted channel coefficients under the spatial split using MLP
%(a) $h_{\mathrm{re}}$ and (b) $h_{\mathrm{im}}$.
%\label{tab:unscaled_g_components}
%}}
%\label{fig:direct_spatial_mlp}
%\end{figure}

A snapshot of the structured data set $\mathscr{D} = \{(P^{Tx,l},P^{Rx,l},g^l_{re},g^l_{im})\}_{l=1}^L$ for a fixed Tx position (m): (91.76, 164.00, 21.44) is shown in Table \ref{tab:unscaled_g_components}. 

\begin{figure}[h]
\centering
\includegraphics[width=\linewidth]{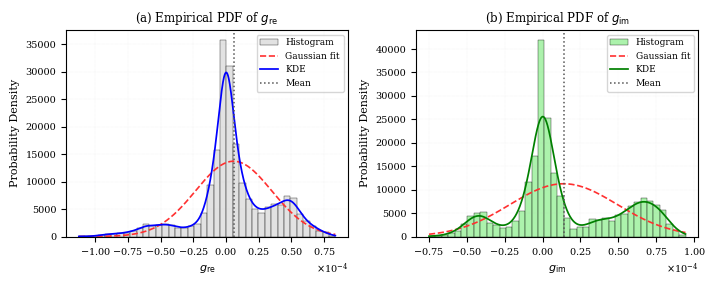}
\caption{PDF Distributions of channel components $g_{\mathrm{re}}$ and $g_{\mathrm{im}}$}
\label{fig:pdf_g_re_and_g_im}
\end{figure}

The probability density functions (PDFs) of real and imaginary components ($g_{re},g_{im}$) of the channel coefficients are given in Fig. \ref{fig:pdf_g_re_and_g_im} and its statistical properties (mean, variance) are given in Table \ref{tab:unscaled_g_statistics}. It is observed that change in shape of pdf plot due to phase compensation.
\begin{table}[htbp]
\centering
\caption{Representative samples of phase-compensated channel components $g_{\mathrm{re}}$ and $g_{\mathrm{im}}$}
\label{tab:unscaled_g_components}
\resizebox{\columnwidth}{!}{%
\begin{tabular}{c c c c}
\hline
\textbf{Sample} &
\textbf{Rx Position (m)} &
$\mathbf{g_{\mathrm{re}}}$ &
$\mathbf{g_{\mathrm{im}}}$ \\
\hline

1 &
$(29.896606,\ 120.701283,\ 1.50)$ &
$2.22065524 \times 10^{-5}$ &
$-6.02497318 \times 10^{-5}$ \\

2 &
$(31.252259,\ 120.697978,\ 1.50)$ &
$-1.14341668 \times 10^{-6}$ &
$-1.53257041 \times 10^{-5}$ \\

3 &
$(32.607911,\ 120.694674,\ 1.50)$ &
$-1.32158149 \times 10^{-5}$ &
$-5.40916525 \times 10^{-5}$ \\

\hline
\end{tabular}%
}
\end{table}

\begin{table}[htbp]
\centering
\caption{Statistical Properties of channel components $g_{\mathrm{re}}$ and $g_{\mathrm{im}}$}
\label{tab:unscaled_g_statistics}

\resizebox{\columnwidth}{!}{%
\begin{tabular}{c c c c}
\hline
\textbf{Component} &
\textbf{Mean} &
\textbf{Variance} &
\textbf{Std. Dev.} \\
\hline

$\mathbf{g_{\mathrm{re}}}$ &
$5.880070 \times 10^{-6}$ &
$8.492258 \times 10^{-10}$ &
$2.914148 \times 10^{-5}$ \\

$\mathbf{g_{\mathrm{im}}}$ &
$1.390088 \times 10^{-5}$ &
$1.255547 \times 10^{-9}$ &
$3.543371 \times 10^{-5}$ \\

\hline
\end{tabular}%
}
\end{table}

\begin{figure}[h]
\centering
\includegraphics[width=\columnwidth]{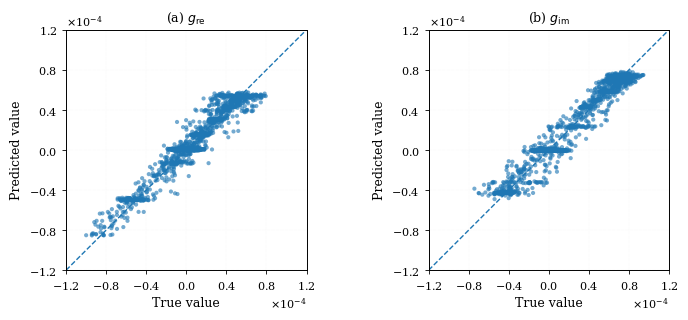}
\caption{
Actual versus predicted channel coefficients under the random split using direct prediction MLP
(a) $h_{\mathrm{re}}$ and (b) $h_{\mathrm{im}}$.
}
\label{fig:phase_random}
\end{figure}

\begin{figure}[h]
\centering
\includegraphics[width=\columnwidth]{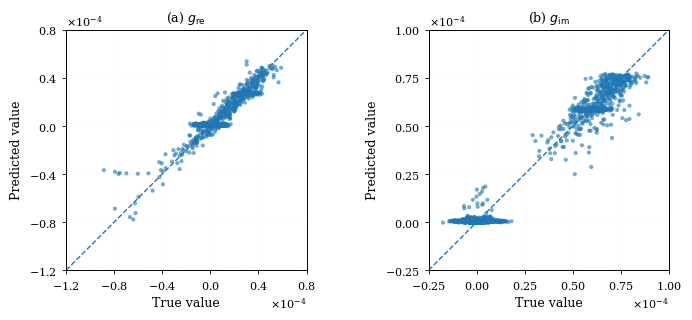}
\caption{
Actual versus predicted channel coefficients under the spatial split using direct prediction MLP
(a) $h_{\mathrm{re}}$ and (b) $h_{\mathrm{im}}$.
}
\label{fig:phase_spatial}
\end{figure}

\begin{figure}[h]
\centering
\includegraphics[width=\columnwidth]{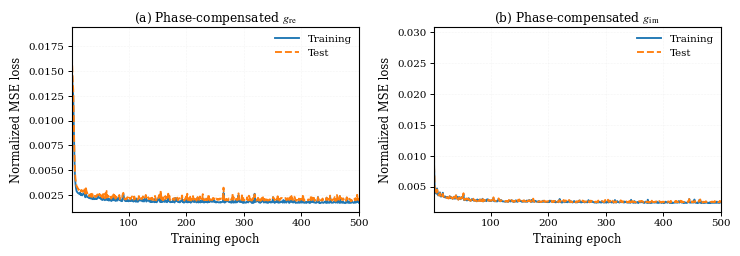}
\caption{
Training and test MSE-loss convergence of the phase-compensated MLP under the random split:
(a) $g_{\mathrm{re}}$ and (b) $g_{\mathrm{im}}$.
}
\label{fig:losscurve_mlp_randomsplit}
\end{figure}

The performance of the trained modles is shown in Table \ref{tab:phase_no_target_scaling}. It is observed that MLP provides better overall performance with $R^2$ of $0.9202$, an adjusted $R^2$ of $0.9199$, an MAE of $0.6269\times10^{-5}$, and an RMSE of $0.8519\times10^{-5}$ under random
split for $g_{re}$. Where as for $g_{im}$ for random split both MLP and XG Boost provides better performance in compared with other regressors. Intrestingly, it is observe that for $g_{im}$ MLP provide much better performance under spatial sampling in compared with random sampling in terms of $R^2$ and adjusted $R^2$ metrics. This might be the samples choosen from group are nearer to the training samples or effect of phase compensation works well for $g_{im}$ in compared with that of $g_{re}$. The performance of the MLP interms of better fit to data is shwon in Fig. \ref{fig:phase_random} and \ref{fig:phase_spatial}. It is observed from the Fig. \ref{fig:phase_spatial} that actual and predictd relationship is not fully diagonal for $g_{im}$. We consider MLP with random split for prediction of $g_{re}$ and $g_{im}$ and corresponding convergence of loss curve is given in Fig. \ref{fig:losscurve_mlp_randomsplit}., 

\begin{table*}[!t]
\centering

\caption{
Model performance for $g_{\mathrm{re}}$ and $g_{\mathrm{im}}$
}
\label{tab:phase_no_target_scaling}

\footnotesize
\renewcommand{\arraystretch}{1.08}
\setlength{\tabcolsep}{4pt}

\begin{tabular*}{\textwidth}
{@{\extracolsep{\fill}}cclcccc@{}}

\toprule

\textbf{Split} &
\textbf{Target} &
\textbf{Model} &
$\mathbf{R^2}$ &
\textbf{Adjusted }$\mathbf{R^2}$ &
\textbf{MAE} &
\textbf{RMSE} \\

\midrule

\multirow{10}{*}{Random}

& \multirow{5}{*}{$g_{\mathrm{re}}$}
& SVR(RBF)
& 0.8681
& 0.8677
& 0.000007533
& 0.000010948 \\

&
& KNN
& 0.9117
& 0.9114
& 0.000006413
& 0.000008959 \\

&
& Random Forest
& 0.9039
& 0.9036
& 0.000006565
& 0.000009348 \\

&
& XGBoost
& 0.9079
& 0.9076
& 0.000006602
& 0.000009150 \\

&
& \textbf{MLP}
& \textbf{0.9202}
& \textbf{0.9199}
& \textbf{0.000006269}
& \textbf{0.000008519} \\

\cmidrule(lr){2-7}

& \multirow{5}{*}{$g_{\mathrm{im}}$}
& SVR(RBF)
& 0.9243
& 0.9240
& 0.000007097
& 0.000009797 \\

&
& KNN
& 0.9400
& 0.9398
& \textbf{0.000006256}
& 0.000008720 \\

&
& Random Forest
& 0.9312
& 0.9310
& 0.000006522
& 0.000009336 \\

&
& XGBoost
& 0.9378
& 0.9376
& 0.000006511
& 0.000008880 \\

&
& \textbf{MLP}
& \textbf{0.9418}
& \textbf{0.9416}
& 0.000006374
& \textbf{0.000008591} \\

\midrule

\multirow{4}{*}{Spatial}

& \multirow{2}{*}{$g_{\mathrm{re}}$}
& XGBoost
& 0.5676
& 0.5662
& 0.000007205
& 0.000010268 \\

&
& \textbf{MLP}
& \textbf{0.8609}
& \textbf{0.8605}
& \textbf{0.000004192}
& \textbf{0.000005824} \\

\cmidrule(lr){2-7}

& \multirow{2}{*}{$g_{\mathrm{im}}$}
& XGBoost
& 0.9166
& 0.9163
& 0.000006209
& 0.000008632 \\

&
& \textbf{MLP}
& \textbf{0.9620}
& \textbf{0.9619}
& \textbf{0.000004235}
& \textbf{0.000005826} \\

\bottomrule

\end{tabular*}

\end{table*}

\begin{figure}[t]
\centering
\includegraphics[width=\columnwidth]{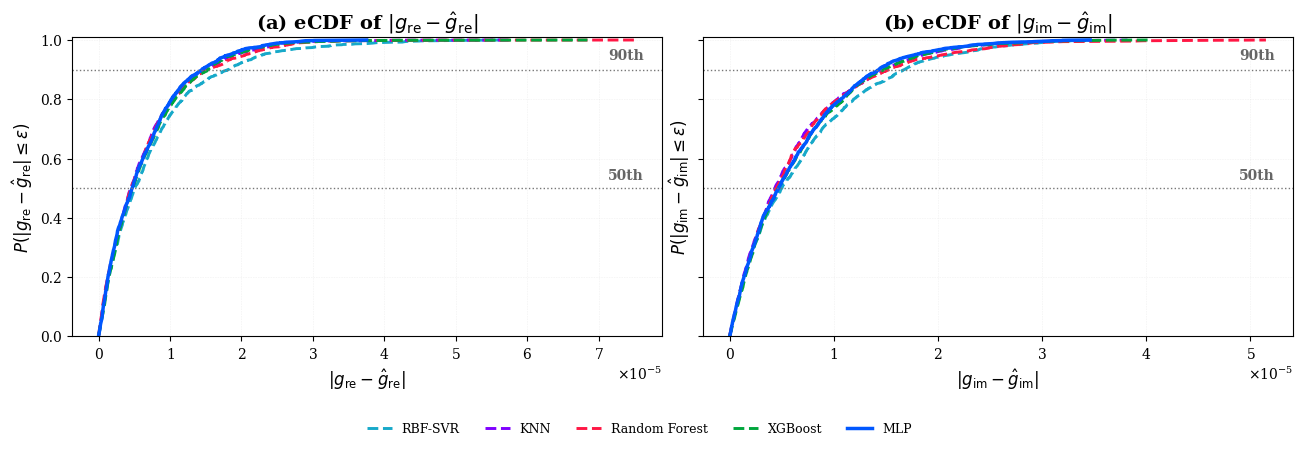}
\caption{eCDF performance comparison for absolute prediction errors of 
$g_{\mathrm{re}}$ and $g_{\mathrm{im}}$}
\label{fig:eCDF_plots}
\end{figure}

An empirical cumulative distribution function (eCDF) which represents the probability of an absolute prediction error less than some threshold $x$ given as $P(|e|\leq x)$. It is shown in Figure~\ref{fig:eCDF_plots} for SVR (RBF), KNN, Random Forest, RGBoost and MLP with random sampling. The eCDF is obtained by first computing the residual error for each test sample as the difference between the true and predicted values of the channel coefficients. The absolute error is then derived from the residuals and sorted in ascending order. The cumulative probability is assigned based on their rank, such that
\begin{equation*}
    F(x)=\frac{1}{N}\sum_{i=1}^{N}\mathbf{1}(|e_i|\leq x)
\end{equation*}
where $N$ denotes the total number of test samples. From the figure, it is observed that $95\%$ threshold indicates that $95\%$ of the prediction errors are smaller than the corresponding error value.The eCDF curves corresponding to XGBoost and MLP exhibits better performance (small error distributions) similar error distributions for both $g_{re}$ and $g_{im}$. A comparison based on the mean absolute error (MAE), summarized in Table~\ref{tab:mean_error_phase_restored}, shows that MLP achieves lower than that of XGBoost model. Also, it is observed that throughout in our computations, performance difference between them is very small.

\begin{table}[htbp]
\centering
\caption{Mean Error Comparison of Phase-Compensated Channel Prediction Models}
\label{tab:mean_error_phase_restored}

\resizebox{\columnwidth}{!}{%
\begin{tabular}{c c c}
\hline
\textbf{Model} &
\textbf{Mean $|e_{\mathrm{real}}|$} &
\textbf{Mean $|e_{\mathrm{imag}}|$} \\
\hline

XGBoost &
$6.602276 \times 10^{-6}$ &
$6.510631 \times 10^{-6}$ \\

MLP &
$\mathbf{6.284505 \times 10^{-6}}$ &
$\mathbf{6.326873 \times 10^{-6}}$ \\

\hline
\end{tabular}%
}
\end{table}

The Table \ref{tab:restored_channel_components}, shows the snapshot of restoration of few samples $\hat{h}_{re}, \hat{h}_{im}$ from predicted $\hat{g}_{re}, \hat{g}_{im}$. It is observed that the error difference between true samples $h_{re}, h_{im}$ and restored stamles are very very small. 

\begin{table}[htbp]
\centering
\caption{restored channel components
$\hat{h}_{\mathrm{re}}$ and $\hat{h}_{\mathrm{im}}$ with their corresponding errors.}
\label{tab:restored_channel_components}
\resizebox{\columnwidth}{!}{%
\begin{tabular}{c c c c c}
\hline
\textbf{Sample} &
\textbf{Rx Position (m)} & \textbf{Channel Components}&
\textbf{Restored $h$} &
\textbf{Error} \\
\hline

\multirow{2}{*}{1} &
\multirow{2}{*}{$(29.896606,\ 120.701283,\ 1.50)$} &
$\hat{h}_{\mathrm{re}}$ &
$-2.21568019 \times 10^{-7}$ &
$-1.826 \times 10^{-21}$ \\

&
&
$\hat{h}_{\mathrm{im}}$ &
$6.42114636 \times 10^{-5}$ &
$0.000 \times 10^{0}$ \\
\hline

\multirow{2}{*}{2} &
\multirow{2}{*}{$(31.252259,\ 120.697978,\ 1.50)$} &
$\hat{h}_{\mathrm{re}}$ &
$1.28577282 \times 10^{-5}$ &
$0.000 \times 10^{0}$ \\

&
&
$\hat{h}_{\mathrm{im}}$ &
$8.41804226 \times 10^{-6}$ &
$0.000 \times 10^{0}$ \\
\hline

\multirow{2}{*}{3} &
\multirow{2}{*}{$(32.607911,\ 120.694674,\ 1.50)$} &
$\hat{h}_{\mathrm{re}}$ &
$3.96517681 \times 10^{-6}$ &
$1.694 \times 10^{-21}$ \\

&
&
$\hat{h}_{\mathrm{im}}$ &
$-5.55413541 \times 10^{-5}$ &
$0.000 \times 10^{0}$ \\

\hline
\end{tabular}%
}
\end{table}

\section{Conclusion}

In this work, we presented a channel estimation method using ML algorithms and MLP at $7GHz$ frequency band. We first generated data set using RT method and used pruning technique to eliminate paths that has insignificant power. We considered only LOS points for data generation. The generated Data set has Tx, Rx locations as features, real and imaginary parts of complex baseband channel coefficients as target variables. Using this data set we trained ML models such as LSVR (RBF), KNN, Random Forest, XGBoost and MLP. We used seperate ML models for real part and imaganery part channel coefficient estimation. Through the simulations/experiments we observed that MLP and XGBoost regression models performs better than remaining models. It is also observed that performance of MLP is better than XGBoost regressor and performance difference between them is very small. One can choose XGBoost if looking for better interpretability or can choose MLP if we look only better performance by compromising on interpretability. Looking forward we are planning to generate real time data and test the same. We are working on improving $R^2$ value by considering image of a particular region as additional input to the model and also extending this work towards multiple antenna use case. 

\section*{Acknowledgment}
The authors would like to thank the International Telecommunication Union (ITU) AI for Good for their valuable meetings and discussions.

\bibliographystyle{IEEEtran}
\bibliography{references}

\end{document}